# An intuitive control space for material appearance

Ana Serrano[1]  Diego Gutierrez[1]  Karol Myszkowski[2]  Hans-Peter Seidel[2]  Belen Masia[1]
[1] Universidad de Zaragoza, I3A   [2] MPI Informatik

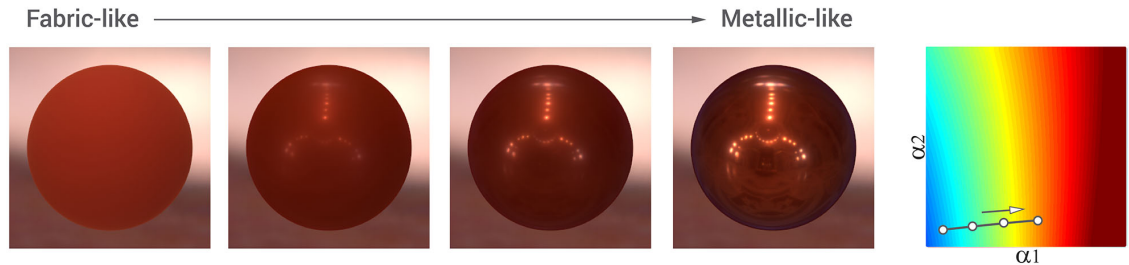

**Figure 1:** *Using our control space to achieve fast, intuitive edits of material appearance. We increasingly modify the metallic appearance of a fabric-like BRDF from the MERL database* (red-fabric2), *yielding intuitive changes in appearance by simply adjusting* one *of our perceptual attributes. Key to this ease of use and predictability of the results is our novel functionals, which map the coefficients of the first five principal components (PC) of the BRDF representation to the expected behavior of the perceptual attributes, based on a large-scale user study comprising 56,000 ratings. The rightmost plot shows the path followed by this edit in our control space. Other applications of our novel space include appearance similarity metrics, mapping perceptual attributes to analytic BRDFs, or guidance for gamut mapping.*

## Abstract

Many different techniques for measuring material appearance have been proposed in the last few years. These have produced large public datasets, which have been used for accurate, data-driven appearance modeling. However, although these datasets have allowed us to reach an unprecedented level of realism in visual appearance, editing the captured data remains a challenge. In this paper, we present an intuitive control space for predictable editing of captured BRDF data, which allows for artistic creation of plausible novel material appearances, bypassing the difficulty of acquiring novel samples. We first synthesize novel materials, extending the existing MERL dataset up to 400 mathematically valid BRDFs. We then design a large-scale experiment, gathering 56,000 subjective ratings on the high-level perceptual attributes that best describe our extended dataset of materials. Using these ratings, we build and train networks of radial basis functions to act as functionals mapping the perceptual attributes to an underlying PCA-based representation of BRDFs. We show that our functionals are excellent predictors of the perceived attributes of appearance. Our control space enables many applications, including intuitive material editing of a wide range of visual properties, guidance for gamut mapping, analysis of the correlation between perceptual attributes, or novel appearance similarity metrics. Moreover, our methodology can be used to derive functionals applicable to classic analytic BRDF representations. We release our code and dataset publicly, in order to support and encourage further research in this direction.

**Concepts:** •**Human-centered computing** → **User studies;** •**Computing methodologies** → **Reflectance modeling; Perception;**

## 1 Introduction

Measurement techniques for material appearance are gaining in accuracy, speed, efficiency, and ease of use (e.g., [Nielsen et al. 2015; Aittala et al. 2015]). This has brought a paradigm shift in computer graphics towards data-driven appearance modeling techniques and databases (e.g., [Matusik et al. 2003; Filip and Vávra 2014; Cornell 2001]). Although this has allowed us to reach an unprecedented level of realism in visual appearance, *editing* the captured data remains a challenge: First, there is a disconnect between the mathematical representation of the data and any meaningful parameters that humans understand; the captured data is machine-friendly, but not human-friendly. Second, the many different formats and representations require handling potentially hundreds of parameters [An et al. 2011; Burley 2012]. And third, real-world appearance functions are usually non-linear and high-dimensional, so editing parameters are rarely intuitive. As a result, visual appearance datasets are increasingly unfit to editing operations, which limits the creative process for scientists, engineers, artists and practitioners in general. In short, there is a gap between the complexity, realism and richness of the captured data, and the flexibility to edit such data.

In this paper, we present a novel intuitive control space suitable for a wealth of applications, such as perceptually-based appearance editing for novice users and non-specialists, developing novel appearance similarity metrics, mapping perceptual attributes to analytic BRDFs, or providing guidance for gamut mapping. Given the existence of large databases of measured BRDFs, a seemingly attractive option would be fitting them to parametric models. Unfortunately, this approach does not suit our goal of flexible material editing well, since the error introduced depends on the nature of the BRDF being represented [Ngan et al. 2005]. Moreover, the error metrics that guide such a fitting do not take into account perceptual aspects, which might lead to visible artifacts for seemingly optimal approximations [Fores et al. 2012]. Last, fitting requires a non-linear optimization which is often numerically unstable, expensive to compute, and typically involves visual inspection to judge the final outcome [Ngan et al. 2005].

Instead, we turn to a non-parametric approach, which can represent with high fidelity a wide scope of measured BRDFs, and lends itself naturally to accommodating our perceptually-based material editing framework. McCool et al. [2001] introduced a log-relative mapping that enables a convenient decomposition of measured BRDFs; later Nielsen and colleagues [2015] performed a linear decomposition into principal components after this mapping. The first five of these components are nicely descriptive of appearance, but cannot be controlled in an intuitive manner. The reason is twofold: First, as the authors discuss, their components are not able to properly isolate the different effects that characterize appearance; and second, as we will show, linear variations in magnitude of these components result in highly non-linear changes in appearance.

We show that there is a much more intricate correlation between principal components, material appearance, and appearance perception. In our work, we first quadruple the original MERL dataset to 400 BRDFs, by synthesizing novel, mathematically-valid samples from measured ones (Sec. 3). We then find a mapping between the space of principal components and higher level perceptual attributes that enable intuitive material editing. This is done as follows: First, we perform a series of experiments to obtain a meaningful list of editing attributes (Sec. 4.1, Exp. 1). From them, a perceptual rating is obtained from a vast user study in which we gather 56,000 answers, covering all our attributes and BRDFs (Sec. 4.2, Exp. 2). We then learn functionals for each of the attributes, mapping the perceptual ratings of each attribute to the underlying principal component basis coefficients (Sec. 5.1). These functionals can be readily used to intuitively and interactively edit measured BRDFs, yielding new, plausible appearances (Sec. 5.2).

We validate the *correctness* of our framework through a user study (Sec. 8) which shows that our functionals can predict well the attribute values given by users. Further, we also show that it is *intuitive* and *predictable*, as well as *versatile*, allowing for a variety of appearance edits; all this can be found in Sec. 8 and the supplemental material. Further, and in addition to editing of measured BRDFs, our derived functionals can be used to increase our knowledge on the perception of appearance (Sec. 6), and for a number of other applications, described in Sec. 7. Finally, we make both our code and dataset public, to foster further research in this direction.

## 2 Related Work

**Editing of parametric models** These works focus mostly on the interface provided to the user. A paradigmatic example of this is BRDF-Shop [Colbert and Pattanaik 2006], where the authors design an artist-friendly editing framework based on an extension of the Ward model. Ngan et al. [2006] propose an image-driven navigation over the space with embedded analytical BRDF models, in which the distance between the models is measured as the difference between rendered images of a sphere under natural illumination. Talton et al. [2009] develop a collaborative editing system that explores the parameter space of the anisotropic Ashikhmin model [2000], based on models saved by other users. Other works focus on fast feedback upon BRDF edits, and treat appearance and lighting jointly [Sun et al. 2007; Cheslack-Postava et al. 2008; Nguyen et al. 2010]. Last, specialized models for car paint design enable BRDF editing by directly specifying the composition of physical paint ingredients, such as density of pigments, or type and distribution of flakes, which affect the appearance of glitter effects [Ershov et al. 2001]. While many of these techniques support measured BRDFs, the common key obstacle is the lack of a sufficiently general and expressive editing space, which we address in this work.

**Editing of non-parametric models** Editing measured BRDF data without fitting to parametric models is a more challenging task, since the editing space is large and unintuitive [Wills et al. 2009]. Lawrence et al. [2006] proposed the Inverse Shade Trees factorization, which decomposes spatially-varying BRDFs into texture and basis BRDFs, which they further decompose into simple 1D curves representing physical effects. Building on their work, Ben-Artzi et al. [2006] proposed a similar framework with precomputed polynomial basis, allowing for complex direct lighting with shadows, as well as interreflections [Ben-Artzi et al. 2008]. All these methods lack intuitive parameters, so that editing implies heuristically modifying a set of 1D curves.

**Industrial standards** A pragmatic approach for a perceptually meaningful characterization of reflectance has been developed by the material industry [Hunter and Harold 1987] and formalized in a number of standardization documents by the American Society for Testing and Materials (ASTM). For example, a number of gloss dimensions have been specified [Wills et al. 2009, Tbl. 1] along with the associated pairs of incident and reflection angles for the reflectance measurements, which should fully characterize the gloss appearance. Westlund and Meyer [2001] derive the correspondence between such isolated reflectance measurements and parameters of selected analytic BRDF models, which effectively links them with the industrial characterization of reflectance in terms of gloss, haze, sheen, and other attributes.

**Perceptual editing spaces** Many different works have applied perceptual strategies in computer graphics [McNamara et al. 2011]. High dimensional perceptual spaces have been used for style similarity [Garces et al. 2014], translucency perception [Gkioulekas et al. 2013], interior design taxonomy [Bell et al. 2013], or shader design [Koyama et al. 2014], to name a few examples. Boyadzhiev et al. [2015] introduce a set of intuitive attributes for image-based material editing. Conceptually, the closest methodology to ours has been proposed for garment simulation, although using the parameters of a custom high-quality production pipeline simulator [Sigal et al. 2015]. For BRDF editing, Pellacini et al. [2000] observed that a direct parameter tuning for analytic BRDFs is often unintuitive due to strongly non-linear changes in material appearance. By analogy to perceptually uniform color spaces such as CIELAB and CIELUV, they derive a perceptually uniform parameter scaling for the Ward model, which has since been used to study image-driven navigation spaces [Ngan et al. 2006], or the influence of shape in material perception [Vangorp et al. 2007]. Wills et al. [2009] extend the concept of perceptually uniform spaces for measured BRDFs, and propose a low-dimensional space suitable for intuitive navigation and construction of new materials, although limited to the achromatic component of reflectance (gloss). Kerr and Pellacini [2010] showed that, for the particular task of matching material appearance, the performance of novice users is comparable for the original Ward model and its perceptually linearized version, while image-driven navigation seems to be less efficient. However, the study is limited to colorless BRDFs, and only for two simple sliders: diffuse and specular.

We draw inspiration from the work of Matusik et al. [2003], who present a data-driven reflectance model. The authors propose to reduce the dimensionality of measured BRDF data either with linear dimensionality reduction (PCA) or with non-linear dimensionality reducers, resulting in a 45D or 15D (respectively) manifold. Then, they define a set of *perceptual traits* (such as redness or silverness), and have a *single* user perform a binary classification whether a given material possesses each particular trait or not. Trait vectors enable navigation in their BRDF spaces by specifying the directions of desirable changes for a given trait or their combinations. Our work is different in many ways: we emphasize on a perceptually meaningful material characterization, but employ a set of carefully selected attributes, which have been identified in a large-scale expe-riment as intuitive, descriptive, and discriminative when describing reflectance properties. We inherit a perceptually meaningful scaling and decomposition of raw BRDF data akin to perceived contrast, which greatly reduces PCA dimensionality [Nielsen et al. 2015], making it comparable to purely perceptually derived spaces [Wills et al. 2009]. We perform dense uniform sampling of the scaled PCA space, synthesizing additional BRDFs from the initial MERL dataset (totaling 400), and obtain ratings for our perceptual attributes in another large scale experiment from which we collect over 56,000 answers from 400 participants. This allows us to reconstruct perceptually-based complex embeddings of our attributes in the PCA-space, which enables intuitive, predictable, and interactive appearance changes from measured BRDF data.

## 3 BRDF representation and database

### 3.1 Principal components space

A database of measured BRDFs can be used to learn a principal components (PC) basis, in which any other BRDF can be represented as [Matusik et al. 2003; Ngan et al. 2006; Nielsen et al. 2015]:

$$\mathbf{b} = \mathbf{Q}\boldsymbol{\alpha} + \boldsymbol{\mu} \qquad (1)$$

where $\mathbf{b} \in \mathbb{R}^N$ is the BRDF represented in the basis, $\mathbf{Q} \in \mathbb{R}^{N \times M}$ is the matrix representing the PC basis (specifically, the eigenvectors of the basis scaled by their eigenvalues), $\mu \in \mathbb{R}^N$ is the average of the measured data, and $\boldsymbol{\alpha} \in \mathbb{R}^M$ are the coefficients of each of the components for the particular BRDF $\mathbf{b}$.

Similar to other approaches [McCool et al. 2001; Nielsen et al. 2015], we perform a log-relative linear mapping of the reflectance data, to avoid allocating most of the available dynamic range to encode variations in the specular peak, and represent our BRDFs with the resulting first five principal components (i.e., $M = 5$), which are loosely related to some characteristics of material appearance [Nielsen et al. 2015]. To make sure that working in a reduced space does not affect the perception of appearance, we have run a small-scale experiment with 20 BRDFs from the MERL database, covering a wide range of appearances. We followed a 2AFC approach, showing the original BRDF, and our representation with only the first five components, and asked the (49) participants to choose which of the two shown images better conveyed a given attribute. The $\chi^2$ analysis of the results (see supplemental material) showed that participants were mostly selecting at random, which indicates that our five dimensional space does not degrade appearance perception and is suitable for our purposes. Examples of the stimuli are shown in Fig. 2, while details on methodology, analysis, and results appear in the supplemental material. Moreover, limiting the space to five dimensions has an additional advantage: When creating a larger database of BRDFs (Sec. 3.2), it helps improve the sampling process by avoiding regions with little impact on appearance.

However, this 5D space, while sufficiently descriptive for our purposes, does not lend itself to intuitive material editing, since the dimensions do not clearly correlate with isolated material properties: Some components (such as the fifth component) are responsible for a combination of different effects, while other effects (such as the shape of the specular highlights) depend on several components. *This suggests that there is a much more intricate correlation between principal components, material appearance, and appearance perception*, as we will show. Moreover, our goal is to be able to modify appearance based on higher-level attributes (such as *glossy*, or *plastic-like*), which do not have a direct, one-to-one correlation with PCA components.

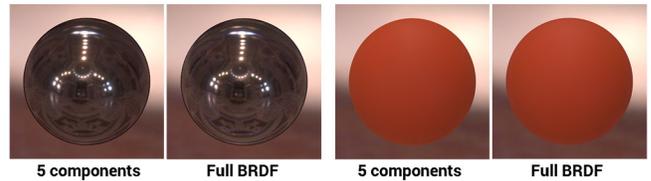

**Figure 2:** *Examples of the stimuli used in our pilot test to determine whether working in a reduced space affects the perception of appearance (shown are ss440 and dark-red-paint, both from the MERL database). The analysis of the results indicates that a five-dimensional space is sufficiently descriptive for our purposes.*

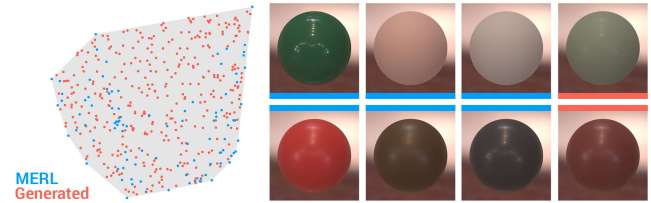

**Figure 3:** Left: *A two-dimensional projection of our 5D BRDF convex hull. Blue dots represent projections of the original MERL BRDFs, while the red dots represent our newly generated samples. Our Gibbs sampling strategy ensures a good coverage of the PC space.* Right: *Each row shows three original BRDFs (blue), plus a novel synthesized material derived from them (red).*

### 3.2 Creating a database of BRDFs

In order to get a sufficiently large amount of data to gather perceptual ratings of our attributes and build the mappings, we need to work with a large BRDF dataset, offering a varied and adequately sampled range of materials with different perceptual properties. Acquiring such a large dataset would be both time consuming and challenging; instead, we opt to synthesize novel BRDFs from existing ones. We choose the MERL database [Matusik et al. 2003] as a starting point, which consists of 100 homogeneous materials that cover reasonably well the range of real world isotropic materials. Similar to their original work, we remove materials with visible non-homogeneities, anisotropy or subsurface scattering, ending up with an initial seed of 94 materials, from which we will synthesize new ones.

It has been shown that the space of mathematically valid BRDFs is convex [Matusik 2003; Wills et al. 2009]. This means that any convex combination of two given BRDFs will produce a new one where non-negativity, energy conservation and reciprocity are preserved. While not all possible combinations will produce a material likely to be found in the real-world, in our work we favor intuitive, *artistic* exploration and expression of material appearance. We therefore compute the convex hull of the measured BRDFs of the MERL database projected in our five-dimensional PCA space, with the goal of synthesizing novel BRDFs inside the polytope defined by the 5D convex hull. When synthesizing our new BRDFs, we aim to achieve a close-to-uniform coverage along our PC dimensions, so that the full space is well represented in our perceptual experiments. Since the exact calculation of a high dimensional polytope is computationally expensive, we choose to approximate a uniform distribution with Gibbs sampling [Metropolis et al. 1953] within the convex hull (see Fig. 3, left).

Then, for each sample, we synthesize a novel BRDF as a convex combination of the three nearest original MERL BRDFs, weighted by their distances to the sampled point. Note that the sampling and the distances are computed in a five-dimensional space, but the convex combination that leads to the novel BRDF is performed on a per-channel basis in a 15-dimensional space (5 x 3 color channels). Fig. 3 (right) shows two novel BRDFs synthesized this way. We generate with this method 306 new BRDFs, yielding a total of 400 different materials for our tests, which can be found in the supplemental material.

# 4 Experiments

We ran a first test to build a user-friendly, intuitive set of attributes for appearance editing; for the sake of conciseness, we only briefly summarize here the main results. In a second test, we obtain a perceptual rating of those attributes, which will allow us to build a mapping between the attributes and the underlying PCA basis coefficients. Please refer to the supplemental material for additional details, including a full description of our first experiment, as well as all the stimuli used.

## 4.1 Experiment 1: Building the space of attributes

For this first test, we rendered a large number of stimuli depicting different materials, built an extensive initial list of candidate appearance descriptors, and then relied on a user study to reduce them to a suitable size. Inspired by recent works on material perception and design (e.g., [Kerr and Pellacini 2010; Jarabo et al. 2014]), our stimuli consist of spheres of 60 different materials from the MERL database [Matusik et al. 2003], chosen to span a wide range of different appearances, and lit by direct illumination. Our initial list was made up of 28 attributes, ranging from high level class descriptors (e.g. ceramic-like) to low level appearance descriptors (e.g., strength of reflections). Relying on Fleming's work [2013], where he states that *we can also make many judgments about the perceived qualities of different materials irrespective of their class membership*, we do not make any restrictions about the type of descriptors in our list. The final list consists of fourteen attributes, covering both high- and mid-level features: *plastic-like, rubber-like, metallic-like, fabric-like, ceramic-like, soft, hard, matte, glossy, bright, rough, tint of reflections, strength of reflections*, and *sharpness of reflections*.

## 4.2 Experiment 2: Measuring the attributes

Once we have built a suitable list of perceptual attributes, our next goal is to characterize a large number of materials based on such a list, which will allow us to derive mappings between attributes and the underlying basis coefficients of the BRDFs. We obtained a total of 56,000 rating responses (400 BRDFs × 10 responses/BRDF × 14 questions/BRDF), which we will use to build the mappings between the perceptual attributes and the underlying PCA coefficients, as described in the next section.

**Stimuli**   To increase the variability of the analyzed BRDFs, we significantly extended our stimuli from the previous experiment, including all our 400 different materials, generated as described in Sec. 3.2. The materials are rendered with PBRT, using the *St. Peter's* environment map. This is also the case for Exp. 1: details on this choice can be found in the supplemental material.

**Participants**   Since we aimed to gather a very large number of answers, we followed similar large-scale studies in computer graphics (e.g., [Rubinstein et al. 2010; Bousseau et al. 2013]) and used Amazon Mechanical Turk[1]. A total of 400 paid subjects took part in our experiment, casting a total of 56,000 rating votes. The feedback we received through the online platform was very positive: they enjoyed the test, and found it engaging and interesting.

[1] Herr and Bostok [2010] recently demonstrated the viability of crowd-sourcing graphical perception studies, reducing variance and finding a good match with results from classic experiments.

**Procedure**   To analyze how different materials are characterized in terms of our list of perceptual attributes, we first considered different options. A valid alternative in principle would be a double-stimulus method, such as a forced-choice pairwise comparison. In such scenario, a ranking (ordering) task could be devised [Parikh and Grauman 2011; Chaudhuri et al. 2013], which is easy for the participants, and usually results in low variance in their responses; however, as Yumer et al. show [2015] a rating approach may be better suited for multi-modal problems like ours, where different BRDFs may have similar attribute strengths. On the other hand, methods to derive a meaningful perceptual scaling from pairwise ranking data exist [Silverstein and Farrell 2001]. Unfortunately, they require close stimuli placement with small attribute differences (ideally overlapping in terms of JNDs), in order to avoid consistent responses where all the votes go to the same stimulus. The lack of a distribution of the user responses might indicate a suprathreshold difference, and does not provide any useful information on attribute scaling. Such a careful placement of the stimuli typically requires extensive pilot studies that would not be practical given the large number of attributes and the 5D embedding that we consider in this work. Another option would be rating pairwise stimuli [Yumer et al. 2015; Koyama et al. 2014]. While this leads to better scaling properties than ranking, it would substantially increase the number of trials, making the tests impractical. Typically a random subset of pairs is considered; only when the parameter space is known, nearby pairs can be selected. (e.g., most of the images lack the rubber-like attribute in our case).

While different pros and cons for each approach can be observed, it has been recently reported after extensive tests that there is no evidence that double stimulus methods are more accurate than single stimulus methods [Mantiuk et al. 2012; Tominaga et al. 2010]. Taking this into account, and in light of the analysis above, we therefore rely on magnitude estimation through rating, also referred to as Mean Opinion Score (MOS). This single-stimulus approach is a well-established methodology, dominant in image and video experiments, and recommended by standard international organizations such as ITU or ISO [ITU 2002; ITU 2008; Keelan 2003].

Similar to previous works [Du et al. 2013; Zell et al. 2015], we chose a five-point scale, which we found offered a good trade-off between the number of options and the difficulty to carry out the test. Each scale was numbered from 1 (*none, or very little*) to 5 (*a lot*). We designed a web-based interface, for easy navigation. The participants' task and the rating scales were explained at the beginning, before proceeding to the actual test. During the test, the participants were shown one rendered material at a time, plus the fourteen perceptual attributes from Exp. 1; they were asked to rate each of the perceptual attributes, for each BRDF, in the Likert-type scale (Fig. S.4 in the supplemental material shows a screenshot of the test). We thus obtained the 56,000 rating responses, which we will use to build the mappings between the perceptual attributes and the underlying PCA coefficients, as described in the next section.

# 5 An intuitive appearance control space

We now describe how we build our mapping between each attribute and the coefficients of the five PCs defining a BRDF, based on the ratings obtained. These mappings will define our intuitive control space for appearance editing.

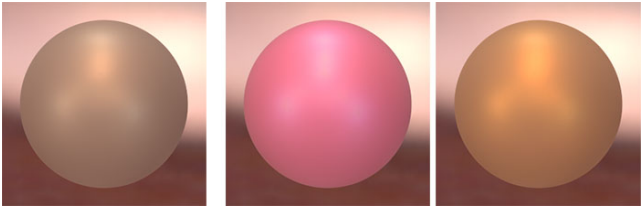

**Figure 4:** *We can modify chromaticity in our framework simply by tuning the chromaticity channels of the CIELab space.* Left: *original BRDF.* Right: *two new BRDFs, generated as described in the text.*

### 5.1 Fitting functionals for the attributes

Similar to related works [Pellacini et al. 2000; Wills et al. 2009], we decouple achromatic reflectance from color information (working in CIELab space), which adds flexibility to our editing framework. Our functionals are derived for achromatic reflectance, but changing chromaticity can be easily accomplished by modifying the $a$ and $b$ channels, as shown in Fig. 4. For each of the attributes, we seek a functional $\varphi : \mathbb{R}^5 \to \mathbb{R}$ that models the behavior of the attribute as a function of the coefficients of the first five PCA components $\boldsymbol{\alpha} = \{\alpha_1..\alpha_5\}$. We will obtain these functions using as input data the ratings given by subjects to each of the BRDFs in Exp. 2, and the $\boldsymbol{\alpha}$ coefficients of those BRDFs. A priori we have no information of how these functions should look like. We therefore choose to use a radial basis function (RBF) network with one hidden layer; RBFs are known to be capable of providing approximations to any continuous function of a finite number of real variables with arbitrary precision [Park and Sandberg 1991], and have been used before to approximate BRDFs [Zickler et al. 2006]. Our RBF network can be expressed as follows:

$$y = \varphi(\boldsymbol{\alpha}) = \sum_{i=1}^{N_c} \theta_i \phi(\|\boldsymbol{\alpha} - \mathbf{c}_i\|) \quad (2)$$

where $y \in \mathbb{R}$ is a value, in the range [0..1], that represents the strength of the attribute, $N_c$ is the number of neurons in the network, $\mathbf{c}_i \in \mathbb{R}^5$ are the centers of such neurons, and $\theta_i$ the weights of each neuron. Following common practice, we model $\phi$ as a Gaussian function, and define the norm as the Euclidean distance:

$$y = \varphi(\boldsymbol{\alpha}) = \sum_{i=1}^{N_c} \theta_i \exp^{-\beta\|\boldsymbol{\alpha}-\mathbf{c}_i\|^2} \quad (3)$$

where $\beta$ controls the smoothness of the Gaussian functions.

For each attribute, we train a network by minimizing the L2-norm between the average attribute values given by subjects for each BRDF (MOS), $y_j \in \mathbb{R}^{N_b}$, where $N_b = 325$ is the number of BRDFs used for training, and the value of $\varphi(\mathbf{x}_j)$. The neuron centers are found by k-means clustering. When choosing the number of neurons $N_c$, we must find a compromise: Too few neurons will not provide enough degrees of freedom to capture the complexity of the space, while having too many has the risk of overfitting the data. We measure the fitting error (MSE) for several values of $N_c$, as well as for neuron-varying values of $\beta$, as shown in Fig. 5 (left), and choose $N_c = 10$ neurons with uniform $\beta$, since larger values of $N_c$ or non-uniform $\beta$ values do not offer a significant decrease in error.

Additionally, we evaluate the goodness-of-fit of the RBFs by calculating, for each attribute and for all the BRDFs in our database, the mean distance between the values predicted by our functionals, and the answers given by each particular user; we plot them by projecting them to a 2D slice ($\alpha_1 - \alpha_2$) of our space. In Fig. 5 (right) we plot these distances for the *rubber-like* attribute (plots for all the attributes are shown in the supplemental material). The low values indicate that our RBFs fit users' opinions well in all regions of the space. Two important conclusions can be derived: First, our RBFs fit the data well in all the regions of the space; second, the high agreement confirms that we can use the MOS as a good approximation of users' opinion. Note that these distances are indirectly indicative of confidence (in the sense of agreement between users): If the variance in users' responses for an attribute and BRDF were high, then the mean distance values plotted here would also be high. Nevertheless, we specifically look into user agreement in Sec. 6.2.

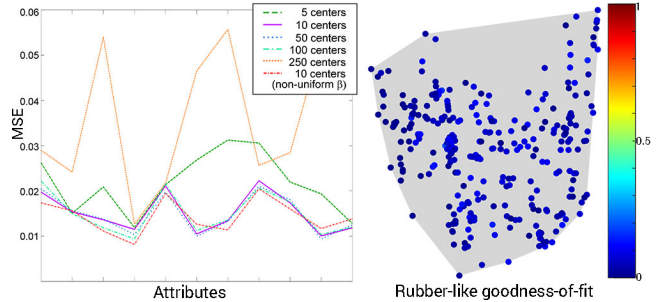

**Figure 5:** Left: *Fitting error (MSE) for our attributes (x-axis). The curves correspond to different numbers of neurons of the RBFs and for non-uniform $\beta$ values (neuron-varying smoothness of the Gaussians); we choose 10 neurons.* Right: *Mean distances between values predicted by our functionals and subjects' answers for the attribute* rubber-like *for each BRDF. The low values indicate that our RBFs fit users' opinions well in all regions of the space, and confirm the adequacy of the use of the MOS in their computation.*

### 5.2 Navigating the space with perceptual attributes

Once we have the functionals mapping a set of coefficients $\alpha_i$ to attribute values $y$, we can interactively navigate our control space. To train our functionals, we map the first five PCA components to each of the attributes individually: $\mathbb{R}^5 \to \mathbb{R}$, obtaining one mapping per attribute. This mapping is surjective (though not bijective). However, for certain applications (e.g., editing) we need the inverse: given an attribute value $y$, find the corresponding coefficients $\alpha_i$. The solution to this inverse problem is not unique; we formulate it as a minimization, which we solve via gradient descent (this will be demonstrated in Sec. 7).

Starting with any BRDF, we compute its $\boldsymbol{\alpha}$ coefficients by projecting it into the PCA basis, and subsequently the values $y_k$ associated to each of the attributes of interest using the mappings $\varphi_k : \mathbb{R}^5 \to \mathbb{R}$, as explained in Sec. 5.1. Given an attribute $k$, and an initial BRDF with coefficients $\boldsymbol{\alpha}_{ini}$, we can alter appearance by modifying its initial value $y_{k,ini}$ to an objective value $y_{k,obj}$. We formulate this as a minimization (we drop $k$ for clarity):

$$\min_{\boldsymbol{\alpha}} \|\varphi(\boldsymbol{\alpha}) - y_{obj}\|^2 \quad (4)$$

where $\varphi(\cdot)$ is given by Eq. 3. We solve this using gradient descent with $\mathbf{x}_0 = \mathbf{x}_{ini}$, aiming to obtain the solution closest to $\boldsymbol{\alpha}_{ini}$ (i.e., the initial BRDF) that satisfies the requirement for $y_{obj}$. To ensure that the solution remains within the five-dimensional convex hull defined by the measured MERL BRDFs, we test at each step whether the new location remains within its boundaries, and stop the minimization if the boundary is reached. Note that, given the correlation between perceptual attributes, changes in one may affect other attributes as well, to reflect a perceptually valid change in material properties. Since our functionals are derived from perceptual ratings given by users, their values will be intuitive and correlate with user perception, as we validate in Sec. 8.

## 6 Analysis and exploration of the space

In this section we first provide a qualitative analysis of our functionals. We then analyze the different attributes, the interactions between them, and the agreement between user responses for different attributes and BRDFs. Finally, we explore the correlation between our attributes.

### 6.1 Qualitative analysis of the attribute functionals

In our work, we map the space of principal components to higher level perceptual attributes that define an intuitive control space for appearance; these mappings will then be used to find the *paths* in PC space that lead to natural-looking appearance changes. Fig. 6 shows a series of 2D slices of our 5D space, defined by the coefficients $\alpha_i$, for different material attributes depicting our mappings using our functionals. We plot two-dimensional slices $\alpha_1 - \alpha_i$ ($i = 2..5$), since the first component $\alpha_1$ has the greatest influence on material appearance. A qualitative analysis reveals interesting insights that align well with our intuition of how we perceive some characteristics of materials. As we explain below, observations on two-dimensional slices of our 5D PC space confirm that: i) analyzing each principal component of the BRDFs in isolation cannot explain how materials are perceived; instead, there are many correlations defined in our larger five-dimensional space; and ii) our approach correlates well with human perception of materials, since we find many expected behaviors in our two-dimensional projections. In the following we describe the different slices in Fig. 6:

- The first slice depicts how the *rubber-like* attribute varies with both $\alpha_1$ and $\alpha_2$ (the specular and diffuse components, respectively). High values of both the specular and diffuse coefficients yield low values for perceived rubber-like, and viceversa. Moreover, as the specular intensity $\alpha_1$ increases, the material becomes less rubber-like, while as the diffuse component increases $\alpha_2$, the material also loses its rubber-like look. This is consistent with our intuition that rubber-looking materials do not show specular highlights and reflect relatively little light overall.

- The second slice analyzes again the $\alpha_1$-$\alpha_2$ plane for *bright*, and shows how both coefficients have an influence on how bright a material looks. Although mainly dominated by $\alpha_2$ (increased brightness is correlated with an increase in the diffuse component), $\alpha_1$ also plays a role: For a fixed value of $\alpha_2$, increasing the specular component also causes the perceived brightness to increase.

- The third slice corresponds to the *metallic-like* attribute, and in this case depicts an $\alpha_1$-$\alpha_3$ (both related to the specular component) cut of the 5D space. For low and mid values of the intensity component $\alpha_1$, the component related to the shape of the specularities $\alpha_3$ plays a significant role: materials appear more metallic as its value decreases. However, for very high values of the intensity, the shape of the specular highlights becomes increasingly irrelevant when identifying the material as metallic.

- In the fourth slice we study again the *metallic-like* attribute, this time as a function of $\alpha_1$ and $\alpha_4$ (Fresnel). As expected, the specular component $\alpha_1$ dominates the metallic look; but we can clearly see an interesting effect: given a value of $\alpha_1$, the perceived metallic quality of the material increases as the Fresnel effect $\alpha_4$ decreases.

- In the last slice, we plot how the *plastic-like* attribute varies with the coefficients $\alpha_1$ and $\alpha_5$. A material is more plastic-like as its specular intensity ($\alpha_1$) increases, as expected; however, the shape of the specular and the Fresnel effect, partially controlled by $\alpha_5$, also play an important role.

### 6.2 Inter-user and intra-cluster agreement

We cluster the measured BRDFs manually into one of six groups according to the *actual* material they belong to, namely *fabric*, *metallic*, *acrylic*, *plastic*, *phenolic*, and *metallic-paint*. We use only measured BRDFs since they can be clustered reliably, following Matusik's naming system [Matusik et al. 2003]. We now seek to analyze the agreement between users when rating each attribute, as well as the agreement between BRDFs from the same material cluster (i.e., whether they share the same appearance).

We obtain, for each cluster and attribute, the *mean score* and a measure of *agreement*. Fig. 7 shows the resulting plots for a sample cluster; the complete plots for all the clusters can be found in the supplemental material. These plots give us a large amount of information about subjective BRDF appearance; in the following, we describe the interpretation of these plots, and present some of the main conclusions.

**Mean score plots** For the mean score we compute the mean value per BRDF per attribute, and box plots showing the interquartile range (IQR, defined as Q3-Q1), and maximum and minimum values (Fig. 7, left). The mean values indicate the general trend of the attribute in the cluster (note that the y-axis is normalized). As with the correlation analysis, the results align with real-world experience; for instance, *metallic-like*, *glossy*, and the *strength* and *sharpness of reflections* all have high mean values for the *metallic* cluster, and much lower for *fabric*. Low variance of the mean for one attribute indicates that such an attribute is a potentially good descriptor of the cluster, while high variance indicates that it is not, since different BRDFs in the cluster are given very different values for such an attribute. For instance, *rough* is not a good descriptor of the *plastic* cluster (Fig. S.6 in the supplemental), which makes sense since plastic materials can have a wide range of surface roughnesses. As a consequence, a consistently high variance of the mean for multiple attributes in a cluster indicates that users do not identify it as a cluster of appearance; this is the case with plastic BRDFs, probably because they can exhibit a wide variety of appearances in the real world. Finally, note that a low IQR (small box plot) in the mean scores indicates that most of the BRDFs in the cluster share the same average value of the attribute, but not necessarily that users agreed when grading such an attribute for each BRDF; instead, it is the agreement box plots that give an indication of user agreement.

**Agreement plots** As a measure of agreement we compute the variance of the scores per BRDF per attribute, and box plots showing the mean of this variance, together with its IQR and maximum and minimum values (Fig. 7, right, and supplemental material). Overall, our plots show consistently low mean values, indicating a large agreement for all clusters and attributes (note that although the maximum value the variance can take is one, the y-axes of the agreement plots range only from 0 to 0.25 for visualization purposes). This suggests that our choice of attributes is adequate for our purposes, being meaningful and intuitive descriptors of appearance; moreover, it also validates using the MOS for the fitting of the functionals. Additionally, a low IQR indicates a good agreement between users for all BRDFs in the cluster, independent of whether the value for the given attribute was high or low (see for instance *glossy* in the *plastic* cluster). A high IQR indicates that for some BRDFs there is agreement, but for others there is not (such as *rubber-like* in the

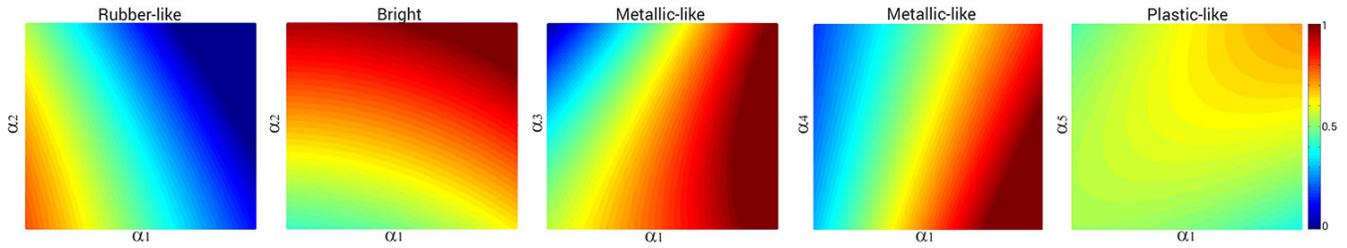

**Figure 6:** *Sample 2D slices of our functionals* $\varphi : \mathbb{R}^5 \to \mathbb{R}$, *mapping coefficients* $\boldsymbol{\alpha}$ *in the PC basis to perceptual ratings for different attributes and along different dimensions. From left to right:* rubber-like *($\alpha_1$-$\alpha_2$ slice);* bright *($\alpha_1$-$\alpha_2$ slice);* metallic-like *($\alpha_1$-$\alpha_3$ slice);* metallic-like *($\alpha_1$-$\alpha_4$ slice); and* plastic-like *($\alpha_1$-$\alpha_5$ slice). Please refer to text for further interpretation.*

*acrylic* cluster, Fig. S.5).

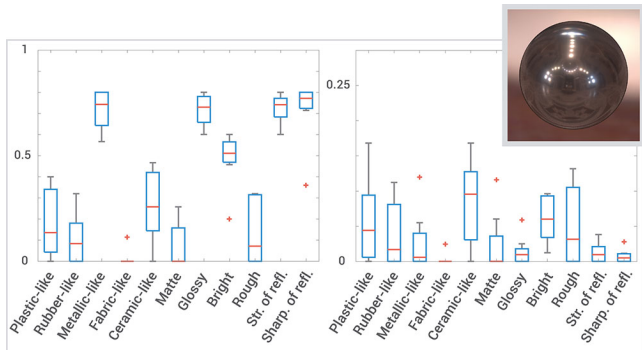

**Figure 7:** *Mean scores (left) and agreement (right) for BRDFs in the* metallic *cluster. The inset shows a sample BRDF from the cluster (*aluminium*). A low mean value in the plot on the right is indicative of a high agreement between users; please see the main text for further explanations.*

**Figure 8:** *Color-coded correlation matrix between attributes (Pearson correlation coefficients): Grey, blue and orange indicate no significant correlation ($p - value > 0.05$), positive, and negative correlations, respectively. Darker shades indicate increasingly stronger correlation, see text for details.*

### 6.3 Correlation between attributes

We finally analyze the *semantic similarity* of our attributes. Fig. 8 shows Pearson correlation coefficients for our attribute set[2]. Grey color indicates no significant correlation ($p - value > 0.05$), while blue and orange indicate positive and negative correlations, respectively. We have additionally highlighted with increasingly stronger shade those pairs showing a larger correlation (three levels, delimited by $> 0.7$ and $> 0.8$ in absolute value). The results match what we would expect: For instance, the *strength* and the *sharpness* of reflections are highly correlated with the perception of how *glossy* a material is, but inversely correlated to how *matte* a surface looks. Similarly, *rough* is highly correlated with *matte*, whereas *rough* and *sharpness* of reflections, or *hard* and *soft*, show a strong negative correlation.

This shows the overall correlation between our functionals, but we can further analyze locally in which particular regions of the space our attributes have a similar behavior: Since our underlying five-dimensional space is of lower dimensionality than the number of attributes, there may be regions in which multiple attributes exhibit the same behavior. In the top part of Fig. 9, we analyze two attributes: *strength of reflections* and *metallic-like*. In each row, we show $\alpha_1 - \alpha_2$ plots for low and high values of $\alpha_3$, $\alpha_4$, and $\alpha_5$. These plots show that both attributes highly depend on $\alpha_1$; this agrees with our intuition, since $\alpha_1$ roughly corresponds to the specular trait of the material. But the plots also reveal that for varying values of $\alpha_3..\alpha_5$, this dependency does not change much for *strength of reflections*, indicating that $\alpha_1$ dominates the value of the attribute in the whole 5D space. The *metallic-like* attribute, however, has a more complex behavior, showing a larger dependency on $\alpha_2$ and $\alpha_3$ in the region of the 5D space defined by low $\alpha_3$ values (see the top-left plot). Despite this difference in this particular region, the overall dependency on $\alpha_1$ translates into the high correlation shown in Fig. 8. In the bottom row of Fig. 9, we further analyze a second pair: *strength of reflections* and *rubber-like*. The latter shows a large dependency on $\alpha_1$ in the regions of the space defined by high values of $\alpha_3..\alpha_5$; this dependency is almost the opposite with respect to *strength of reflections*, as one would expect (the stronger the specular, the less rubber-like it appears). In addition, *rubber-like* shows a very different behavior in the regions where $\alpha_3..\alpha_5$ have low values, indicating that $\alpha_2$, related to the diffuse component, plays a crucial role in our perception of rubbery appearance in certain cases. In particular, for low values of $\alpha_3$, indicative of a high roughness, the relative importance of $\alpha_1$ decreases in favor of $\alpha_2$; this is also the case for the metallic-like attribute explained before.

Figs. 8 and 9 also show that our different perceptual attributes are not orthogonal. This was already observed by Matusik et al. [2003], while a similar conclusion was reached for perceptual parameters designed for garment simulation [Sigal et al. 2015]. This is a desirable characteristic for a control space, since it prevents the user from trying to produce a BRDF that is *glossy* and *matte* at the same time, for instance.

---
[2] We obtained similar results in terms of trends and significance using Spearman rank correlation; they can be found in the supplemental.

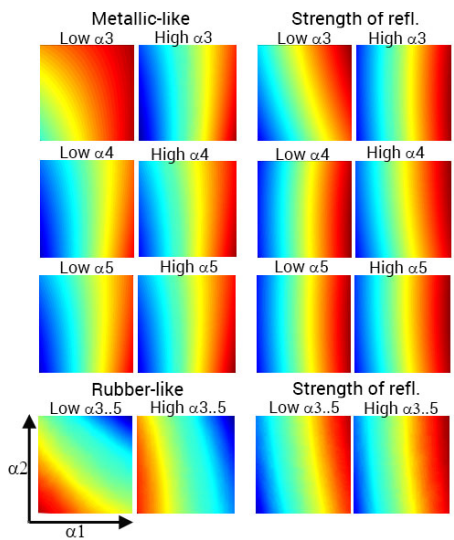

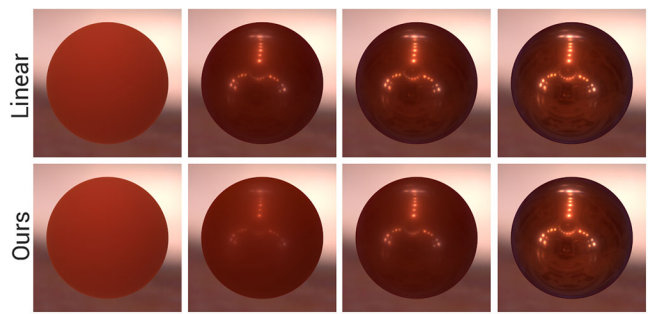

**Figure 9:** *Comparison of the behavior of three attributes in different regions of our five-dimensional space. Attributes are: (i)* metallic-like*, (ii)* strength of reflections*, and (iii)* rubber-like*. (i) and (ii) exhibit a high positive correlation, so they behave similarly in most regions of the space. On the contrary, (ii) and (iii) exhibit a high negative correlation, behaving very differently in most regions. Colormaps depict the value of the corresponding attribute in the $\alpha_1 - \alpha_2$ plane of our 5D space for different values (low/high) of the remaining coefficients ($\alpha_3..\alpha_5$).*

**Figure 10:** *Comparison between linear interpolation in the log-relative mapped PC-space of Nielsen et al. [2015] and traversing our perceptually-based space, going from a fabric-like to a metallic-like BRDF. Our edits are more perceptually-uniform, whereas a linear interpolation in PC-space causes sudden, unpredictable changes in appearance.*

## 7 Applications

**Material editing** As we have seen, our mappings reveal complex relationships between PC components and appearance attributes, so intuitive edits cannot be performed directly on the PC components. Moreover, since our mappings are derived from perceptual ratings, changes in the attributes are more predictable and intuitive than changes in the PC coefficients. We illustrate this in Fig. 10: the top row shows a linear interpolation from a fabric-like BRDF to a metallic-like BRDF in the log-relative mapped PC-space of Nielsen et al. [2015]; the bottom row shows such a transition using our perceptual attributes. Our transitions look more equally spaced in terms of appearance, while moving linearly in the PC-space yields sudden and non-linear changes in appearance. Fig. 11 shows more edited BRDFs using our framework, obtained from measured BRDFs from the MERL database. For each original BRDF, we linearly vary the value of one of our perceptual attributes, and render the resulting BRDF at each step. The resulting BRDFs are obtained using the procedure described in Sec. 5.2. As the figure shows, our user-friendly editing space yields feasible and appealing edited BRDFs, while keeping variations perceptually meaningful. The last row shows the path followed through our 5D space while varying each of the attributes (we show the most representative 2D slice). Additionally, in Fig. 12 we make significant changes in the appearance of the teapots, for input BRDFs very different in nature. All the edits have been achieved by tuning *a single attribute* in our control space. More edited BRDFs with different illuminations can be found in the supplemental material.

**Similarity metrics** Similarity metrics are a useful tool to determine if two images are visually equivalent, i.e., if they convey the same impressions of scene appearance [Ramanarayanan et al. 2007]. Establishing a measure of similarity between two BRDFs would be very useful for a large number of applications, including gamut mapping, BRDF compression, fitting, or even acquisition. Different metrics have been proposed, such as root mean squared error (RMSE) and its variants (cosine-weighted, with or without cubic root) [Fores et al. 2012], or the perceptually uniform reparametrizations of analytic BRDF spaces [Pellacini et al. 2000]. However, the definition of a global similarity metric does not allow to analyze perceptual attributes *separately*: Two BRDFs could have very similar specular peaks, yet strikingly different diffuse properties. Our functionals offer a novel approach, providing a means for evaluating similarity for *individual visual attributes*. Fig. 13 shows an example with three BRDFs (A, B, and C). Pairwise comparisons (A-B, and A-C) using an RMSE similarity metric [Fores et al. 2012] yield very similar results (6358 vs. 6365), although it seems obvious to think of A-B as more similar than A-C. Instead, our functionals allow to break down the notion of similarity in terms of specific aspects of appearance. For instance, in terms of brightness, our metric accurately yields a much closer distance between A-B (0.0452), than A-C (0.5171).

**Perceptual attributes for analytic BRDFs** Our methodology, together with the subjective data compiled in our user studies, can be used to derive novel functionals relating our perceptual attributes to other BRDF representations, such as analytic models. We fit our database (the 400 BRDFs) to a chosen model, and then use the answers collected in our user study to train new functionals relating the set of perceptual attributes to the parameters of the analytic model. Fig. 14 demonstrates this for the *blue-fabric* BRDF from MERL's database, fitted to a microfacet model with a Beckmann distribution. The middle image shows the learned functional for the chosen attribute. We can see how the attribute's value varies non-linearly with the parameters $k_s$ (specular) and $A$ (roughness) of the model: For low roughness the influence of $k_s$ in the strength of reflections is much larger, as one would expect. Our functional hides the complexity of the parameters' interactions in the original representation, making the editing process more predictable and intuitive. Since our extended database will be made public, this will facilitate the learning of new functionals for any other BRDF representation, as well as the exploration of the resulting spaces.

**Guidance for gamut mapping** Gamut mapping is a classic problem in appearance modeling: The goal is to ensure a good correspondence between the original model and its reproduction, overcoming the limitations of the output medium (e.g., 2D/3D printing). Since our functionals allow to obtain different BRDFs with the same perceived apperance in terms of a given attribute, we can use them to guide out-of-gamut cases back into the set of representable appearances, while ensuring that the desired *attribute* is not changed. This is shown in Fig. 15, for two pairs of BRDFs along isocontours of the functional $\varphi(\boldsymbol{\alpha})$, and two different attributes. Note that two BRDFs located along an isocontour of an attribute should keep the same appearance in terms of that attribute, but may have different reflectance properties overall. Our functionals therefore expand the range of gamut mapping strategies beyond classic approaches.

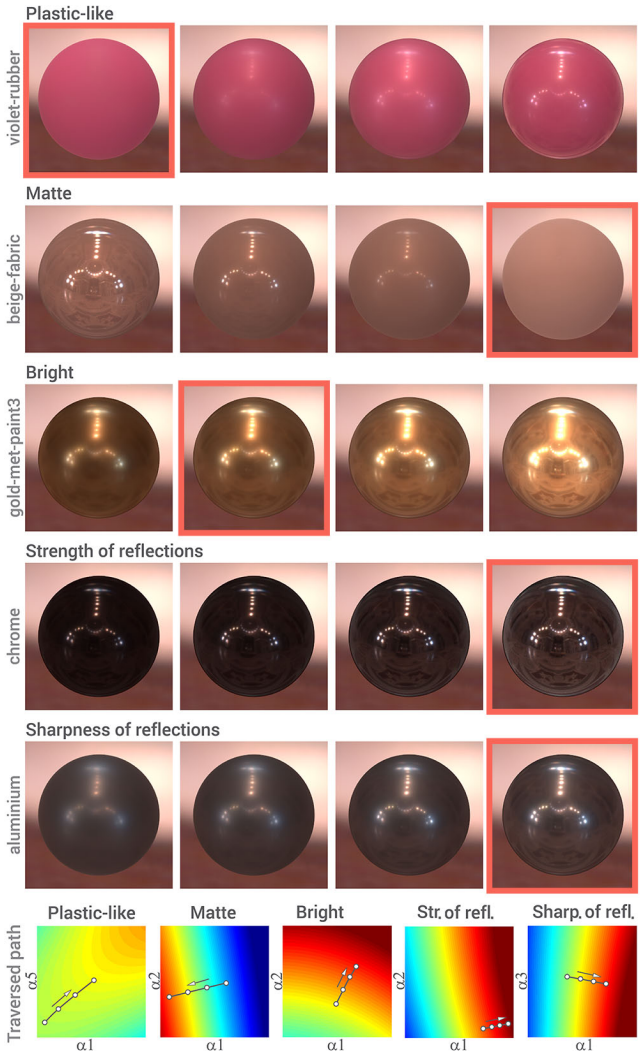

**Figure 11:** *Editing BRDFs by varying the values of our attributes, using our functionals trained for the different attributes. Each row shows an original BRDF from the MERL database (marked in orange), and the results when linearly increasing or decreasing the value of the given attribute. The last row shows the path followed when traversing our 5D space (we show the most representative 2D slice) to compute each row.*

## 8 Validation

**Predictability** When a user edits a BRDF by tuning our perceptually-based attributes, she would need to know what to predict from each adjustment. Our functionals allow this (as we have shown in Fig. 10), facilitating this desired predictability. Nevertheless, here we set out to further validate this, by verifying whether our functionals can really predict the magnitude of an attribute that is perceived by users. We design a user study, following the same procedure as in Exp. 2 (Sec. 4). We chose 44 BRDFs, both measured and edited, and asked 60 new participants to rate our fourteen perceptual attributes for each one. Each participant had to rate ten BRDFs. To analyze the results, we first average the ratings for each attribute and BRDF. Then we compute, for each attribute, the MSE across BRDFs between the subjective ratings and the values predicted by our functionals. We obtain a very accurate match, with MSE values below 0.02 for most of the attributes, and not getting higher than 0.03 for any of them. Tbl. 1 shows the exact values for the fourteen attributes. This close match clearly indicates that our functionals are excellent predictors of perceived appearance.

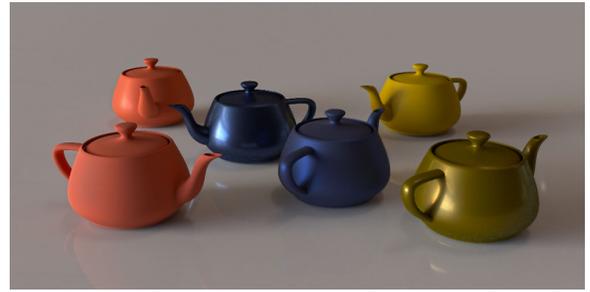

**Figure 12:** *Scene rendered with measured materials from the MERL database and edited BRDFs using our framework. From left to right, we modify MERL's* pink-plastic *BRDF (foreground) by increasing the* ceramic-like *attribute (background); we modify MERL's* blue-metallic-paint2 *BRDF (background) by increasing the* rough *attribute (foreground); we modify MERL's* yellow-paint *BRDF (background) by increasing the* metallic-like *attribute (foreground).*

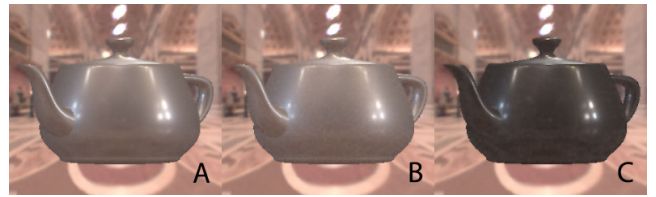

**Figure 13:** *Our functionals allow to define similarity metrics based on individual appearance attributes. While A-B and A-C are very similar according to RMSE, A-B exhibit a much lower distance in terms of brightness than A-C.*

**Proof of concept with novice users** To assess the practicality of our control space for novice users, we designed an informal user study involving appearance editing. The task is to be carried out using both our functionals and the commercial software 3ds Max from Autodesk. We created a prototype implementation of our approach as a plugin for BRDF Explorer[3]. Users were shown images of two spheres, rendered with an initial and a final BRDF. Equivalent pairs were created with each software (our BRDF plugin and 3ds Max) to guarantee fairness, since an exact appearance match is difficult to achieve across platforms, and not a requisite for the test. The pairs were chosen so that they covered a wide range of appearances, showing complex and significant changes between the two (shown in the supplemental material). Users were then asked to create spheres with an intermediate appearance, using both platforms in succession. Editing with both tools produced similar results in terms of appearance, but much faster with our prototype, as we show in Tbl. 2. The resulting BRDFs appear in the supplemental material.

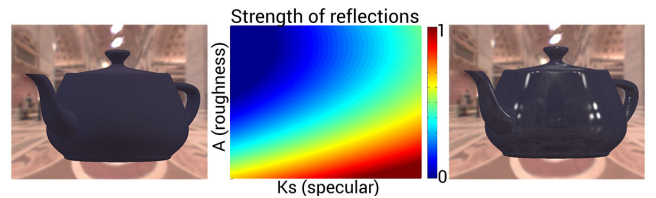

**Figure 14:** *Editing the blue-fabric BRDF from MERL's database, fitted to a microfacet model (Beckmann distribution). We learn a novel functional that hides the complexity of the interactions of the different parameters in the original representation, and use it to easily alter a particular attribute. From left to right: original BRDF fitted to the microfacet model, learned functional for strength of reflections, and edited BRDF (increasing strength of reflections).*

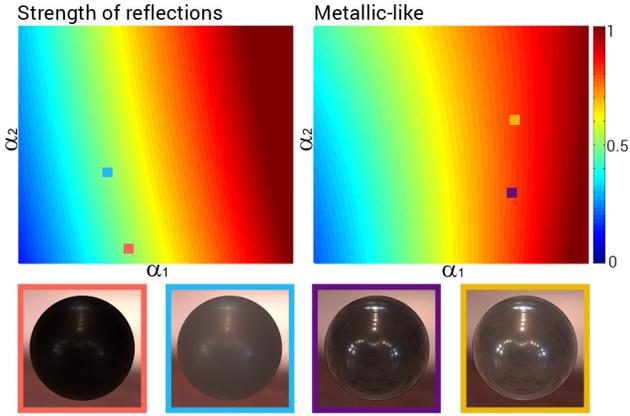

**Figure 15:** *2D slices on the $\alpha_1$-$\alpha_2$ plane for two attributes:* strength of reflections *and* metallic-like. *On each slice, we render the BRDFs corresponding to two points along an isocontour of the respective attribute. Two BRDFs located along an isocontour of an attribute should have the same appearance in terms of that attribute, but may have different reflectance properties overall.* Left: *the strength of the reflections is kept the same, despite varying roughness and diffuse properties.* Right: *the metallic quality of both BRDFs is very similar, despite having different strength and sharpness of reflections.*

**Table 1:** *MSE between the subjective ratings and the values predicted by our functionals, for each attribute, showing how our functionals are excellent predictors of perceived appearance. Error is in the range [0..1].*

| Attribute    | MSE    | Attribute          | MSE    |
|--------------|--------|--------------------|--------|
| Plastic-like | 0.0062 | Matte              | 0.0125 |
| Rubber-like  | 0.0127 | Glossy             | 0.0178 |
| Metallic-like| 0.0127 | Bright             | 0.0195 |
| Fabric-like  | 0.0161 | Rough              | 0.0170 |
| Ceramic-like | 0.0133 | Strength of refl.  | 0.0134 |
| Soft         | 0.0224 | Sharpness of refl. | 0.0214 |
| Hard         | 0.0271 | Tint of reflections| 0.0149 |

The outcome of this user study suggests that novice users find it challenging to convey a particular appearance, and that our editor can be useful in such cases. In conclusion, while commercial packages like 3ds Max offer a very good degree of control with many advanced features, for novice users our system offers a more intuitive and easy-to-use control. Both approaches are thus complementary; our system could be integrated as a plugin to these more sophisticated commercial packages, extending the range of tools available for predictable appearance changes.

---
[3]http://www.disneyanimation.com/technology/brdf.html

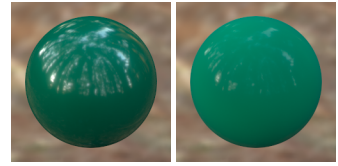

**Figure 16:** *Limitation example. Since our dataset does not contain a large number of fabric-like samples, editing towards that goal from a very different initial BRDF may fail. In this case, our system cannot remove the specularities present in the original BRDF.*

**Table 2:** *Editing times (in seconds) with the commercial tool 3ds Max and with our prototype for each of the tasks. Note that the table includes the times actually spent editing (i.e., rendering times employed on intermediate visualization of results are subtracted from the total measured times).*

|        |         |        | Time (in seconds) | |
|--------|---------|--------|---------|---------------|
|        |         |        | 3ds Max | Our prototype |
|        |         | User A | 41      | 20            |
|        | Pair #1 | User B | 194     | 9             |
|        |         | User C | 94      | 35            |
|        |         | User A | 95      | 40            |
| Task 1 | Pair #2 | User B | 108     | 66            |
|        |         | User C | 39      | 35            |
|        |         | User A | 76      | 38            |
|        | Pair #3 | User B | 80      | 43            |
|        |         | User C | 91      | 29            |

## 9 Discussion and conclusion

We have presented an intuitive control space for material appearance, which allows for artistic exploration of plausible material appearances based on perceptually-meaningful attributes. We have significantly extended the original MERL database to include 400 BRDFs, both captured and synthesized. We have derived novel functionals connecting principal components of the BRDF to a high-level characterization of material appearance, inferred from 56,000 answers collected in a large-scale study with 400 participants. This characterization is made up of our appearance attributes, which are intuitive, descriptive, and discriminative with respect to many different reflectance properties, as we have shown. We have further analyzed the resulting appearance space, which has yielded insights on material perception, and proposed a number of example applications that can benefit from our approach. Our framework aligns changes of the attribute values with predictable appearance changes. Similar to related works [Matusik et al. 2003; Sigal et al. 2015], our attributes are not orthogonal; this is to be expected, and we have shown that indeed some appearance characteristics are highly correlated.

There are many opportunities for interesting future work. First, we do not claim to have found a complete, universal list of perceptual attributes defining appearance. This is an open problem, for which no established methodology exists. In fact, a key advantage of our flexible methodology is that it allows to define *custom* attributes, which may adapt better to a particular user or context, while avoiding mixed nomenclatures. Moreover, it can also be used on different databases. Second, it would be interesting to expand our approach to more materials; despite the fact that our extended MERL dataset provides a reasonably uniform coverage over a very wide range of isotropic appearances, some perceptual attributes can be under-represented. This translates into less user ratings, which may lead to less reliable functionals in some regions of our 5D space (see Fig. 16). Similarly, for some BRDF clusters and specific attributes we find that the variance in scores is relatively high (e.g., *ceramic-like* for the *metallic* cluster in Fig. 7). This seems to indicate that subjects do not agree on how *ceramic-like* the BRDFs in that cluster are. Our functionals will thus be less reliable in that case, as a consequence of people not agreeing on perceptual appearance. Large variance in scores for an attribute in a cluster, however, can have different causes: Some attributes have a large variance in scores, but a high agreement (e.g., *rough* for the *metallic* cluster in Fig. 7), seemingly indicating that the large variance in scores comes from the fact that that particular attribute can exhibit a range of different values within the cluster; in the case of the metallic cluster, BRDFs show a wide range of roughness. Our data, however, is not enough to state strong conclusions in this regard. Further, our system does not currently handle some complex appearance behavior such as color changes, grazing angle effects, or hazy gloss. These are undersampled in our dataset, and remain as future work, deserving further investigation. Last, despite the many insights gained in this work, a full exploration of our space for material appearance still remains an exciting open task.

We hope that our work can inspire additional research, in addition to the four applications we have shown. For instance, it could help to better understand the underlying perceptual aspects of analytic models, or to find a perceptual scaling for their parameters (Fig. 14 shows a proof of concept mapping between perceptual attributes and analytic BRDFs). It could also help to examine the representational space of existing models, to design computational fabrication techniques to achieve a desired appearance, or even to develop efficient BRDF sampling strategies.

## Acknowledgements

We thank the members of the Graphics & Imaging Lab for fruitful insights and discussion, and in particular Elena Garces, Adrian Jarabo, Carlos Aliaga, Alba Samanes, and Cristina Tirado. We would also like to thank Oleksandr Sotnychenko. This research has been funded by the European Research Council (Consolidator Grant, project Chameleon, ref. 682080), as well as the Spanish Ministry of Economy and Competitiveness (project LIGHTSLICE, ref. TIN2013-41857-P). Belen Masia and Ana Serrano would like to acknowledge the support of the Max Planck Center for Visual Computing and Communication. Ana Serrano was additionally supported by an FPI grant from the Spanish Ministry of Economy and Competitiveness.